# Random Walk Model on a Hyper-Spherical Lattice

S. Boettcher [a] [*]

[a]Department of Physics, Brookhaven National Laboratory,
Upton, New York 11973, USA

We use a one-dimensional random walk on $D$-dimensional hyper-spheres to determine the critical behavior of statistical systems in hyper-spherical geometries. First, we demonstrate the properties of such walk by studying the phase diagram of a percolation problem. We find a line of second and first order phase transitions separated by a tricritical point. Then, we analyze the adsorption-desorption transition for a polymer growing near the attractive boundary of a cylindrical cell membrane. We find that the fraction of adsorbed monomers on the boundary vanishes exponentially when the adsorption energy decreases towards its critical value. We observe a crossover phenomenon to an area of linear growth at energies of the order of the inverse cell radius.

The study of random walk models in statistical mechanics has led to a variety of practical applications[1] and analytical results.[2] They provide insights into the nature of critical phenomena observed in more complicated systems.

In this letter we propose a new model of random walks that substantially simplifies the study of statistical systems in arbitrary dimensions. As an example of the subtle effects that our method reveals for statistical systems near curved boundaries, we discuss the adsorption-desorption transition of polymers growing near an attractive boundary.[3] In the limit of an infinitely extended polymer, a finite fraction $P(\kappa)$ of monomers gets adsorbed on the boundary as soon as the attractive potential $\kappa$ on the boundary increases above a critical value. For planar boundaries one generically finds that $P(\kappa)$ vanishes linearly when $\kappa$ approaches $\kappa_c$ in the adsorbed phase.[4] With our model we can study such a system near a curved boundary like a cylindrical cell membrane. Here, a larger entropy weakens the transition. For a cylinder of radius $m \geq 0$ and for $\Delta\kappa \equiv \kappa - \kappa_c \to 0_+$, we find

$$P(\kappa) \sim \frac{4}{81} \frac{e^{-\frac{8}{9(m+1)\Delta\kappa}}}{(m+1)\Delta\kappa^2}. \qquad (1)$$

This asymptotic expression ceases to be valid when $\Delta\kappa \sim 0.188/(m+1)$ where we observe a crossover to planar behavior.

---

[*]This work was done in collaboration with M. Moshe at the Physics Department of the Technion in Haifa, Israel.

Consider an infinite set of concentric and equally spaced spheres in arbitrary spatial dimension $D$. $S_n$, $n = 0, 1, 2, 3, \ldots$, designates the surface of the $n$th sphere from the center with area $2\pi^{D/2} n^{D-1}/\Gamma(D/2)$. We define a random walk on such a configuration in the following way:[5] Let $c_{t,n}$, $t \geq 0$, $n \geq 1$, be the probability for a random walker to be located *anywhere* between $S_{n-1}$ and $S_n$ at time step $t$. The walker may have the choice to walk outward with a (normed) probability proportional to the total surface area of the sphere just outward, $P_{out}(n) = n^{D-1}/\mathcal{N}(n)$, and similarly inward with $P_{in}(n) = (n-1)^{D-1}/\mathcal{N}(n)$. Thus, we find a $1+1$-dimensional evolution equation to describe the behavior of the walker in this $D$-dimensional geometry:

$$c_{n,t} = P_{out}(n-1)\, c_{n-1,t-1} + P_{in}(n+1)\, c_{n+1,t-1}. \quad (2)$$

For the case of concentric spheres it was found that the probability of ever returning to the origin for a random walker starting at the origin is given by $\Pi_D = 1 - 1/\zeta(D-1)$ for $D > 2$, and unity for $D \leq 2$. This result is in qualitative agreement with $\Pi_D$ for a hyper-cubic lattice. In Ref. 5, it has been shown that one-dimensional random walks on a hyper-spherical lattice have the same scaling behavior as $D$-dimensional random walks on a hyper-cubic lattice.

These ideas can be extend to the following di-rected percolation problem: For $t = 0$, there is one "wet" site on an infinitely extended line of "dry" sites with unit spacing. At even times $t$, sites take on only integer values, at odd times $t$, sites take on only half-integer values, $\pm 1/2$, $\pm 3/2$, .... If at time $t-1$ two neighboring sites at $i$ and $i+1$ are "dry", then at time $t$ the site at $i+1/2$ is also "dry". Furthermore, we assume that two "wet" sites at $i$ and $i+1$ at time $t-1$ *always* produce a "wet" site at $i+1/2$ at time $t$. The latter assumption makes the percolation cluster, i. e. the region of all "wet" sites at any time $t$, compact and there are only two interfaces between the compact "wet" cluster in the middle and the two surrounding "dry" regions. Thus, we merely have to specify the following three situations: On the next time step the gap between the interfaces either makes a unit step outward with a probability $P_{out}$, or it makes a unit step inward with a probability $P_{in}$, or both interfaces shift to the right or to the left without widening the gap with a total probability $P_{stay}$. The behavior of such a compact percolation cluster can be mapped into the one-dimensional random walk similar to Eq. (2) by defining $c_{n,t}$ to be the probability that the gap between both interfaces at time $t \geq 0$ is of width $n \geq 0$. Generally, we find the percolation probability

$$P = \left[ \sum_{n=1}^{\infty} \prod_{i=1}^{n-1} \frac{P_{in}(i)}{P_{out}(i)} \right]^{-1}, \qquad (3)$$

independent of $P_{stay}$. For a percolating $D$-bubble, $P_{out}(n) = q^2(n+1)^{D-1}/\mathcal{N}(n)$ and $P_{in}(n) = (1-q)^2 n^{D-1}/\mathcal{N}(n)$, $n \geq 1$, with $P_{out}(0) = 1$, we obtain $D$-dependent critical coefficients and we compute the percolation probability

$$P(q, D) = \left[ \sum_{n=0}^{\infty} \left( \frac{1-q}{q} \right)^{2n} (n+1)^{1-D} \right]^{-1} \qquad (4)$$

For $D = 1$ we reproduce the result given in Ref. [6]. In Fig. 1 we plot Eq. (4). For all $D$ we obtain $q_c = 1/2$. For $D < 2$ the transition to percolation is second order, first order for $D > 2$, and there is a tricritical point at $D = 2$.

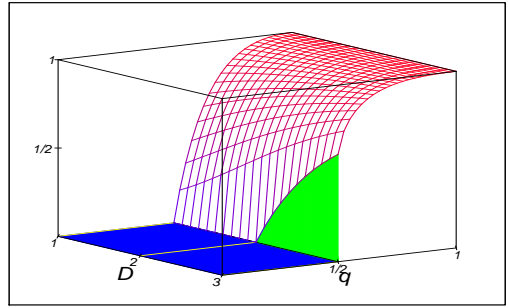

Figure 1. Percolation probability $P(q, D)$ for directed compact percolation in a curved geometry.

As an application of these ideas, we discuss the adsorption fraction for an extended polymer growing near an attractive cylindrical boundary.[7] Consider a lattice consisting of an infinite set of concentric cylinders of unit spacing. Let the innermost cylinder – the surface of the boundary – be of integer radius $m \geq 0$, the next innermost of radius $m+1$, and so on. Each cylinder is labeled by its radius. Now consider a random walk in unit steps either parallel to the length or perpendicular to these cylinders, starting on the boundary. A parallel step is taken with a relative weight of 1, steps outward and inward are taken with relative weights of $P_{out}(n) = 2(n+1)/(2n+1)$ and $P_{in}(n) = 2n/(2n+1)$, respectively, for $n > m$, while on the boundary $P_{out}(m) = 1$, $P_{in}(m) = 0$. Modeling a growing polymer, at each step the walk attains a fugacity $z$, while for each step on the boundary it also acquires a potential energy of $\kappa \geq 1$.

A walk with $L > 0$ parallel steps has reached $L$ levels, $\{h_i\}_{i=1}^{L+1}$, $h_i \geq m$, above or on the boundary. We want to restrict these walks such that $|h_{i+1} - h_i| \leq 1$, $0 \leq i \leq L$. The transfer matrix $T_{h_{i+1}, h_i}$ that describes the transition of the walker from the $i$th to the $(i+1)$st level is

$$\begin{aligned} T_{j,i} &= z^{|j-i|} \kappa^{\delta_{m,j}} \\ &\quad [\delta_{j,i} + P_{out}(i)\delta_{j-1,i} + P_{in}(i)\delta_{j+1,i}]. \end{aligned} \qquad (5)$$

The partition function of all walks is found to be

$$Z(z, \kappa) = z \vec{b}^{(t)} T^2 (1 - zT)^{-1} \vec{e}. \qquad (6)$$

If $\lambda_{max}$ is the largest eigenvalue of $T$, then $Z$ diverges for $z \nearrow z_\infty(\kappa) = 1/\lambda_{max}$. The fraction of adsorbed monomers $P(\kappa)$ in the limit of an infinitely long polymer is given by

$$P(\kappa) = -\frac{\kappa}{z_\infty(\kappa)} \frac{dz_\infty(\kappa)}{d\kappa}. \quad (7)$$

To obtain a non-vanishing adsorption fraction it is necessary that the attractive potential $\kappa$ is larger than some critical value, $\kappa_c$. In terms of the eigenvalues $\lambda = \lambda(z, \kappa)$ of the transfer matrix $T$, $\kappa_c$ is found to be the smallest value of $\kappa$ for which $T$ has a bound state on the line $z = z_\infty(\kappa)$.

The spectrum of the transfer matrix is determined by the eigenvalue problem $\lambda g_n = \sum_{i=m+1}^\infty T_{n,i} g_i$. With $\epsilon = 2z/(\lambda - 1)$ and $\gamma = (1 - \sqrt{1 - \epsilon^2})/\epsilon$, we find the eigenvalue condition

$$1 = \frac{\gamma \left[(m+1)(2z+\epsilon)(\kappa - 1) - z\kappa\right] F_{m+1}}{z\kappa\epsilon(m + 1/2)F_m}, \quad (8)$$

where $F_m = F(1/2, m; m + 1/2; \gamma^2)$ is a hypergeometric function. We obtain $z_\infty(\kappa)$ implicitly from Eq. (8) by replacing $\lambda = 1/z_\infty(\kappa)$. Bound states are found for $\lambda \geq 1 + 2z$, or equivalently $\kappa \geq \kappa^*(z) = (1 + 2z)/(1 + z)$. We obtain $\kappa_c = 4/3$, $z_c = 1/2$ at the intersection of $\kappa^*(z)$ and $z_\infty(\kappa)$. Inserting $z_\infty(\kappa) = z_c - \Delta z(\kappa)$ and $\kappa = \kappa_c + \Delta\kappa$ into the equation for $z_\infty$, we get in the limit $\Delta z \to 0_+$, $\Delta\kappa \to 0_+$:

$$\Delta z(\kappa) \sim \frac{1}{48} e^{-\frac{8}{9(m+1)\Delta\kappa}}, \quad \Delta\kappa \ll 1/(m+1), \quad (9)$$

neglecting exponentially smaller corrections. Using Eq. (7), we arrive at Eq. (1). In Fig. 2 we plot the exact values of $P(\kappa)$ for $m = 0$, 1, 2, 3 and $m = \infty$ for $\kappa \geq \kappa_c = 4/3$.

The simplicity of the dynamics in this new random walk model, and the nontrivial scaling obtained from it, raise interesting questions regarding the universal properties of this lattice. At the critical transition only a few fundamental properties of the system determine its behavior. In this model, the critical behavior arises from the balance between a short-range attractive potential and the spatial entropy. We argue that these features are sufficiently well represented by a random walk on a hyper-spherical lattice. It has been shown that such a lattice reproduces all the universal scaling properties expected of lattices.[5] The critical behavior obtained on this lattice for rotationally symmetric systems should therefore reflect the universal critical behavior of the system. The advantage of random walks on hyperspheres is to describe the critical behavior in a minimal and tractable way in comparison, for example, to a more structured hyper-cubic lattice.

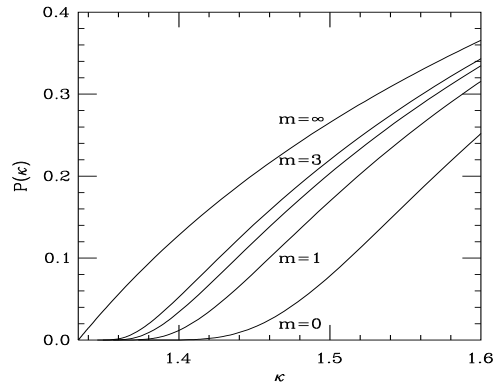

Figure 1. The exact adsorption fraction $P(\kappa)$.